\documentclass[sigconf]{acmart}
\newcommand\fnurl[1]{%
  \footnote{\url{#1}}%
}
\usepackage{subcaption}
\usepackage{multirow}
\usepackage[table,usenames]{}

\usepackage{amsmath}

\usepackage{booktabs} 
\usepackage{mathtools}
\usepackage{balance}
\usepackage{booktabs}
\newcommand{\ra}[1]{\renewcommand{\arraystretch}{#1}}

\copyrightyear{2018} 
\acmYear{2018} 
\setcopyright{acmlicensed}
\acmConference[JCDL'18]{The 18th ACM/IEEE Joint Conference on Digital Libraries}{June 3--7, 2018}{Fort Worth, TX, USA}
\acmPrice{15.00}
\acmDOI{10.1145/3197026.3197038}
\acmISBN{978-1-4503-5178-2/18/06} 

\begin{document}
\title[Entropy-Based Topic Modeling on Domain-Specific Collections]
{My Approach = Your Apparatus?}
\subtitle{Entropy-Based Topic Modeling on Multiple Domain-Specific Text Collections}


\author{Julian Risch}
\affiliation{%
  \institution{Hasso Plattner Institute, University of Potsdam}
  \streetaddress{Prof.-Dr.-Helmert-Str. 2--3}
  \city{14482 Potsdam} 
  \state{Germany} 
}
\email{julian.risch@hpi.de}

\author{Ralf Krestel}
\affiliation{%
  \institution{Hasso Plattner Institute, University of Potsdam}
  \streetaddress{Prof.-Dr.-Helmert-Str. 2--3}
  \city{14482 Potsdam} 
  \state{Germany} 
}
\email{ralf.krestel@hpi.de}

\renewcommand{\shortauthors}{J. Risch et al.}

\begin{abstract}
Comparative text mining extends from genre analysis and political bias detection to the revelation of cultural and geographic differences, through to the search for prior art across patents and scientific papers.
These applications use cross-collection topic modeling for the exploration, clustering, and comparison of large sets of documents, such as digital libraries.
However, topic modeling on documents from different collections is challenging because of domain-specific vocabulary.

We present a cross-collection topic model combined with automatic domain term extraction and phrase segmentation. 
This model distinguishes collection-specific and collection-independent words based on information entropy and reveals commonalities and differences of multiple text collections.
We evaluate our model on patents, scientific papers, newspaper articles, forum posts, and Wikipedia articles. 
In comparison to state-of-the-art cross-collection topic modeling, our model achieves up to $13\%$ higher topic coherence, up to $4\%$ lower perplexity, and up to $31\%$ higher document classification accuracy. 
More importantly, our approach is the first topic model that ensures disjunct general and specific word distributions, resulting in clear-cut topic representations.
\end{abstract}

%
%





\maketitle

\section{Cross-Collection Topic Models}
\label{sec:introduction}
A variety of information retrieval and data mining applications deals with unstructured text data.
With unsupervised machine learning, topic models iteratively estimate probabilistic representations of topics and documents.
Based on word co-occurrence frequencies, topic models cluster text documents by their latent topics and thus structure large document collections.
Despite the overwhelming, unstructured amount of data, topic models thereby enable users to search for documents by topic, to explore document collections, and to extract meaningful information.

As digital libraries grow, getting an overview and keeping track of large document collections becomes even more important.
However, combining the knowledge from multiple collections is challenging.
Linguistic contrasts, such as domain-specific vocabulary, complicate topic modeling.
Cross-collection topic models extend previous single-col\-lec\-tion models to multiple collections.
They aim to model document-topic representations despite linguistic contrasts and to reveal per-topic similarities and differences of collections.

An exemplary application of cross-collection topic modeling is the search for prior art during the patent examination process~\cite{risch2017what}.
For this task, related work from any publicly available text collection, such as granted patents or scientific papers, needs to be retrieved.
Patent-specific words, such as ``device'' or ``apparatus'' and paper-specific words, such as ``algorithm'' or ``approach'', hamper the effective usage of keyword-based approaches and word co-occurrence statistics.

Another field of application is bias detection in newspapers.
For this task, we consider each newspaper's articles as an individual collection.
We distinguish collection-independent and collection-specific words based on their frequency distribution: 
Collection-independent words have similar frequency across all collections, whereas collection-specific words have significantly different frequency per collection.
Still, collection-specific words might occur in all collections, but they occur much more frequently in one particular collection compared to other collections.
In newspapers, col\-lec\-tion-specific words serve as a bias indicator, because these words occur more frequently in one newspaper.
Further bias analysis then focuses on interpretation of these differences in language use.
Similarly, cross-collection topic models can identify linguistic differences in document collections with different cultural and regional background.
For example, topic modeling on traveler forum posts about different countries aims to reveal country-specific words per topic and regional and cultural differences in forum posts.

As opposed to previous work, we focus on domain-specific language and consider collections as domains with specific vocabulary.
We propose a novel topic model that combines state-of-the-art cross-collection topic modeling~\cite{paul2009crosscultural} with the concept of termhood from the field of automatic domain term extraction~\cite{chang2005domain}.
With the help of an entropy-based termhood measure, our model ranks words according to their collection-specificity.
To split the vocabulary in a set of collection-specific words and collection-independent words, the model sets an entropy threshold and estimates the proportion of collection-specific words for each dataset individually and automatically.
As a consequence, the topic model guarantees a clear-cut separation of words with collection-specific and collection-independent frequency distribution.
A mixture of collection-specific and collection-independent word distributions represents each latent topic.
The precise distinction of collection-specific and collection-independent words is a novelty in cross-collection topic modeling.
Furthermore, in order to resolve semantic ambiguity of single words, our topic model considers also multi-word phrases.
We evaluate our model and a state-of-the-art approach with regard to three quality measures: topic coherence, language model perplexity, and document classification accuracy.
The evaluation is based on four datasets with either two or three collections.

Section~\ref{sec:related_work} presents related work in the fields of probabilistic topic models and automatic domain term extraction. 
We present our topic model together with the entropy-based termhood estimation, the entropy threshold definition, and the estimation procedure of the proportion of collection-specific words in Section~\ref{sec:approach}. 
Section~\ref{sec:evaluation} describes the experiment setup and compares the proposed model with state-of-the-art cross-collection topic modeling. 
We conclude with a summarization of our contributions and paths for future work in Section~\ref{sec:conclusion}.

\section{Related Work}
\label{sec:related_work}
\paragraph{Latent Dirichlet Allocation}
\label{sec:relatedlda}
Being the standard topic model for single text collections, latent Dirichlet allocation (LDA) models each document as a probability distribution over topics and each topic as a probability distribution over words~\cite{blei2003latent}.
Gibbs sampling~\cite{griffiths2004finding} is used for the estimation of these latent distributions, as it is less complex to implement and achieves up to two orders of magnitude faster runtime compared to variational Bayes and expectation propagation~\cite{yao2009efficient}.
While LDA is based on the assumption that documents with similar topic distribution exhibit similar word distribution, documents from different collections use collection-specific words and thus have different word distributions.
As a consequence, LDA is not suited for modeling topics on multiple collections with domain-specific vocabulary.

\paragraph{Cross-Collection Latent Dirichlet Allocation} 
\label{sec:relatedcrosscollection}
Comparative text mining extends text mining techniques to more than one document collection for the purpose of revealing collection's similarities and differences.
The cross-collection mixture model (ccMix) discovers common topics across collections of news articles from different publishers and of product reviews from different companies~\cite{zhai2004cross}.
To extend topic modeling to multiple collections, ccMix draws each word either from a topic's collection-independent or a topic's respective collection-specific word distribution.
However, ccMix uses only a single, user-defined parameter as the probability that words are collection-in\-de\-pen\-dent or collection-specific. 
Therefore, ccMix cannot distinguish collection-specific and collection-independent words precisely.
Instead of a single, user-defined parameter, cross-collection LDA (ccLDA) learns a probability distribution of collection-independent and collection-specific words per topic and per collection~\cite{paul2009crosscultural}.
Applied to traveler forum posts about the UK, Singapore, and India, ccLDA identifies topics about ``food'' or ``weather'' alongside with collection-specific (region-specific) words per topic, such as ``masala'' and ``seafood'' or ``monsoon'' and ``snow''.
An alternative to collection-specific topic distributions is proposed by Eisenstein et al.: Their approach models a background word distribution and the differences in log-frequencies from this distribution for collection-specific words~\cite{eisenstein2011sparse}.

More recently, supervised cross-collection topic models as extensions of LDA and ccLDA have been proposed for cross-domain learning~\cite{bao2013partially,gao2012supervised}.
Interestingly, no topic model so far utilizes a word's frequency of occurrence across all collections in order to determine whether the word is collection-specific or not.
They all lack the ability to precisely identify a word as collection-in\-de\-pen\-dent if it occurs with similar frequency of occurrence across all collections.
Counterintuitively, in ccMix, ccLDA, and subsequent models, a word can be simultaneously part of a collection-in\-de\-pen\-dent and a collection-specific word distribution.
In contrast, our approach guarantees a clear-cut distinction of collection-specific and col\-lec\-tion-in\-de\-pen\-dent words.

\paragraph{Multi-Lingual and Other Cross-Collection Topic Models}
\label{sec:relatedmulti}
Multi-lingual topic modeling deals with stronger linguistic differences between document collections.
Zhang et al.\ model topics across different languages by incorporating a bilingual dictionary into a probabilistic topic model~\cite{zhang2010cross}.
While their approach requires a dictionary, our experiments on multi-lingual Wikipedia articles demonstrate that we are able to model multi-lingual topics without a dictionary.
LDA can be extended to multiple collections by running several LDA instances in parallel, unified by manually identifying common topics~\cite{zheng2014comparative} or with a hierarchical model of per-topic word distributions~\cite{chen2015differential}. 
On a newspaper dataset with different regions as collections, the hierarchical model learns a master topic across all regions and region-specific descendant topics.
However, the model does not distinguish collection-specific from collection-independent words and descendant topics across different regions are almost identical for generic topics.
In contrast, our approach guarantees for each topic that the collection-independent topic representation has no word in common with each collection-specific topic representation.
Another contribution of Zhang et al.\ focuses on asymmetric and weakly-related collections, extending ccLDA with collection-specific topics~\cite{zhang2015fast}. 
In contrast, we focus on collection-specific words in common topics.

Fang et al.\ identify verbs, adjectives, and adverbs as differing opinion words in three news collections per topic~\cite{fang2012mining}.
They model only one word distribution per topic and assume that different collections share many words per topic and that only opinion words differ.
This assumption does not hold for collections with strong linguistic contrasts and especially not for multi-lingual datasets.
Furthermore, their model does not distinguish collection-specific and collection-independent opinion words.
To the best of our knowledge, no previous work models topics on patents and scientific papers simultaneously.
However, there is related work that models topics of these collections separately, such as a topic-model-based recommender system for prior art in patents~\cite{krestel2013recommending} and a topic model on abstracts of scientific papers~\cite{griffiths2004finding}.

\paragraph{Automatic Term Extraction}
\label{sec:relateddomainspecific}
The field of automatic (domain) term extraction (ATE) deals with the extraction of words or compounds of multiple words that are considered domain-specific terms in text documents.
To this end, Kageura and Umino define termhood as ``the degree that a linguistic unit is related to domain-specific concepts''~\cite{kageura1996methods}.
Several papers address the extraction of domain-specific terms by identifying term candidates based on part-of-speech patterns and ranking them by their termhood afterwards.
It is a common assumption that: ``The information that a term candidate carries is also an important indicator of its termhood.''~\cite{kit2002corpus}.
Therefore, term candidates are ranked according to their termhood, which is measured with variants of term frequency-inverse document frequency (TF-IDF).

In contrast to TF-IDF, Inter-Domain Entropy (IDE) considers the distribution of a word's relative term frequency across all domains~\cite{chang2005domain}.
The closer this distribution is to a uniform distribution, the lower is a word's termhood.
While IDE provides a ranking of words according to their termhood, the ranked list needs to be split into domain-specific and domain-independent word sets based on a threshold, which varies for different datasets.
However, most previous work arbitrarily considers the top-10\% or top-100 words to be specific terms or sets thresholds empirically~\cite{chang2005domain,fedorenko2013automatic}.
Instead of statistical information about word frequencies, Li et al.\ incorporate semantic information from learned latent topics into a novel termhood measure~\cite{li2013novel}.
On single collections, Wilson and Chew extend the standard LDA model with different term weighting schemes~\cite{wilson2010term}. 
Thereby, the dominance and scattering of high-frequency words can be controlled and stop word removal becomes obsolete.
In our work, we combine topic modeling and domain term extraction to model multiple document collections and their linguistic characteristics.

\paragraph{Topic Models on Phrases}
In order to reduce semantic ambiguity, text mining applications can take multi-word phrases into account (instead of single words only).
For example, the unigrams ``support'', ``vector'', and ``machine'' have a different meaning if they are processed as a phrase and therefore topic models based on both unigrams and bigrams outperform unigram topic models at information retrieval tasks.~\cite{wang2007topical}.
Topical phrase mining has been applied to model medical terms~\cite{He:2016:ETP:3016100.3016316}, combined with topical change over time~\cite{Jameel:2013:UTS:2484028.2484062}, and scaled to large datasets~\cite{El-Kishky:2014:STP:2735508.2735519}.
With a supervised approach, Kawamae et al.\ investigate the relationship between training labels and corresponding phrases~\cite{Kawamae:2014:SNT:2556195.2559895}.
While phrases can be modeled as a hierarchy of Pitman-Yor processes~\cite{Lindsey:2012:PTM:2390948.2390975}, Lau et al.\ show that also n-gram tokenization as a pre-processing step improves topic quality~\cite{Lau:2013:CTM:2483969.2483972}.
Recently, a data-driven approach automates phrase segmentation with robust performance at different domains~\cite{liu2015mining,shang2017automated}.
Hence, topic modeling and phrase segmentation can be applied in independent, automatic steps.
We include automatic phrase segmentation in our proposed topic model during the pre-processing step of tokenization.

\section{Entropy-Based Topic Modeling}
\label{sec:approach}
Our approach combines ccLDA as a cross-collection topic model with an entropy-based measure of termhood.
As a result, we propose a novel topic model that splits the vocabulary in collection-specific and collection-independent words and provides more meaningful topic mixture representations of documents.
Collection-specific and collection-independent words form each topic's representation and reveal commonalities and differences of the collections.

\subsection{Basic Cross-Collection Model}
Our model and  ccLDA have in common that they both contain per-topic collection-independent word distributions $\varphi$ and per-topic and per-collection collection-specific word distributions $\sigma$.
Furthermore, there are per-document topic distributions $\theta$.
These three distributions have Dirichlet priors $\alpha$, $\beta$, and $\delta$.
Which word $w$ is sampled in a document depends on the document's collection $c$, the topic $z$ of word $w$, and the binary decision variable $x$.
$x$ determines whether $w$ is sampled from a collection-independent or a collection-specific word distribution.
The original ccLDA samples $x$ from a Bernoulli distribution for each occurrence of a word.
As a consequence, out of two different occurrences of the same word, one occurrence might be considered collection-specific and the other collection-independent.

\subsection{Extended Cross-Collection Model}
The main difference compared to ccLDA is the following: ccLDA determines separately for each occurrence of a word whether this occurrence in particular is collection-specific, whereas our model determines globally whether a word is collection-specific for all its occurrences across all documents.
If our model assumes that a word is collection-specific, this assumption holds for every document in which the word occurs.
In contrast, if ccLDA assumes that a word is collection-specific, another occurrence of the same word in the same sentence could be considered collection-independent.
Furthermore, ccLDA learns a per-topic and per-collection probability distribution from which the model randomly samples whether a particular occurrence of a word is collection-specific.
In contrast, our model does not sample this property randomly but uses a precise differentiation of termhood on the vocabulary level.

According to an entropy-based termhood measure, the vocabulary is split into two sets: collection-specific and collection-in\-de\-pen\-dent words.
With regard to this measure, $x$ is sampled from a Bernoulli distribution $\psi$ with the parameter $\gamma$.
$\gamma$ corresponds to the proportion of occurrences of collection-specific words in the entire dataset and is described together with the entropy-based termhood measure in Section~\ref{sec:entropy}. 
Via the Bernoulli distribution $\psi$ and the proportion of collection-specific words $\gamma$, the entropy-based termhood measure determines the mixture of words from $\varphi$ and $\sigma$ across all documents.
Section~\ref{sec:evaluation} gives examples of how this conceptual difference causes significant change in resulting topics.
Figure~\ref{fig:entropy-basedmodel} depicts the graphical model of our approach and the corresponding generative process is as follows:
\begin{enumerate}
     \item Draw a collection-independent multinomial word distribution $\varphi_{z}$ from Dirichlet($\beta$) for each topic $z$.
     \item Draw a collection-specific multinomial word distribution $\sigma_{z,c}$ from Dirichlet($\delta$) for each topic $z$ and each collection $c$.
     \item For each document $d$, choose a collection $c$ and draw a topic distribution $\theta_d$ from Dirichlet($\alpha$).
     \begin{enumerate}
     \item For each word $w$ in $d$, draw a topic $z$ from $\theta_d$.
     \item Draw $x$ from Bernoulli distribution $\psi$ according to proportion of collection-specific words $\gamma$.
     \item If $x = 0$, draw $w$ from $\varphi_{z}$; else from $\sigma_{z,c}$\footnote{The conceptual difference in distinction from ccLDA is to draw $w$ from \emph{disjoint} $\varphi_{z}$ and $\sigma_{z,c}$ according to \emph{entropy-based} $\gamma$.}.
      \end{enumerate}
  \end{enumerate}

\begin{footnotesize}
\begin{figure}[!htbp]%
\begin{minipage}[]{.49\columnwidth}%
\centering%
\includegraphics[width=0.9\columnwidth]{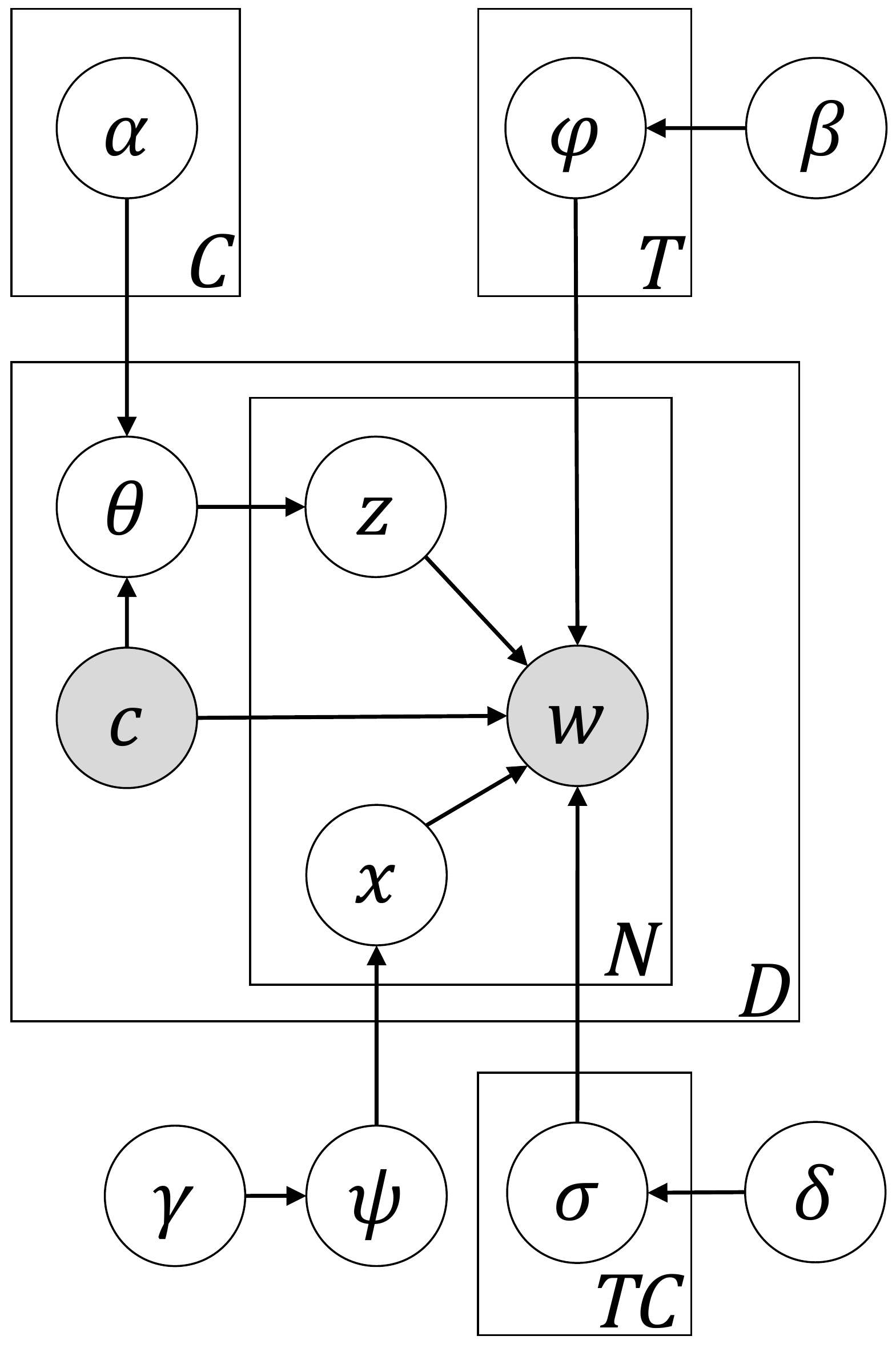}%
\end{minipage}%
\begin{minipage}[]{.49\columnwidth}%
\centering%
\begin{tabular}{|ll|}%
\hline%
$c$ &collection\\
$w$ &word\\
$x$ &determines whether $w$ \\
&depends on $\sigma$ or $\varphi$\\
$z$ &topic\\
$C$ &number of collections\\
$T$ &number of topics\\
$D$ &number of documents\\
$N$ &number of words/doc.\\
$\alpha$, $\beta$, $\delta$ & Dirichlet priors\\
$\gamma$ &proportion of \\
&collection-specific words\\
$\theta$ &topic distribution\\
$\sigma$ &collection-specific \\
&word distribution\\
$\varphi$ &collection-independent \\
&word distribution\\ 
$\psi$ &Bernoulli distribution\\
& of variable $x$\\ 
\hline%
\end{tabular}%
\end{minipage}%
\caption{Graphical representation of the proposed entropy-based topic model.}
\label{fig:entropy-basedmodel}%
\end{figure}
\end{footnotesize}

\subsection{Cross-Collection Word Entropy}
\label{sec:entropy}
In this paper, we distinguish collection-specific and collection-independent words based on entropy.
For each word, the entropy of its frequency distribution across all collections measures the word's termhood. 
Intuitively, the termhood shall be low for words that are evenly distributed across documents from all collections.
In contrast, the termhood shall be high for words that occur more frequently in documents of a specific collection.
Together with an entropy threshold, the termhood determines whether a word is collection-specific.
In our case, a random variable $X$ represents the collection $c$ of a given word $w$, where $P(X=c) = P(c|w)$ holds. 
In the following formulas, we consider a word $w$ as an entry in the vocabulary and not as a particular occurrence of a word in a document. 
Thus, we define the normalized entropy $H(w)$ as:
\[
H(w) = \frac{1}{log_2 C}\displaystyle\sum_{c=1}^{C} -P(c|w)\cdot log_2 P(c|w).
\]
If $w$ is uniformly distributed across all collections, $P(c|w)$ is equal for all $c\in[1,C]$ and $\sum_{c=1}^{C} -P(c|w)\cdot log_2 P(c|w)$ reaches its maximum, $log_2 C$.
To obtain an entropy value in the interval of [0,1] that reaches its maximum for collection-specific words, we normalize entropy with the factor $log_2 C$.
We consider $P(c|w)$ as the probability that the document's collection is $c$, given the occurrence of word $w$. 
Because the probability $P(c|w)$ is unknown a priori, we estimate this frequency on training documents before the topic sampling process.

\subsection{Estimation of Word Probabilities}
In order to estimate posterior $P(c|w)$, we estimate evidence $P(w)$, prior $P(c)$, and likelihood $P(w|c)$ on training documents: 
\begin{itemize}
 \item $P(w)$ is the probability that word $w$ is randomly chosen from any document in the entire dataset.
 \item $P(c)$ is the probability that a randomly chosen word is from a document of collection $c$.
 \item $P(w|c)$ is the probability that word $w$ occurs in a randomly chosen document of collection $c$.  
\end{itemize}

Due to the inherent sparsity of limited training data, some words occur only once in the entire corpus.
With only a single observation of such word, called hapax legomena, there is only limited knowledge about this word's true frequency distribution across collections.
Although hapax legomena have been never observed in any other collection, they might have a non-zero probability of occurrence.
For this reason, we use Laplace smoothing in order to create pseudocounts for unobserved words. 

Based on entropy and the corresponding termhood\footnote{Termhood of word $w$ corresponds to $1-H(w)$}, we can sort words according to their estimated collection-independence.
To incorporate this sorted vocabulary into the topic model, we define an entropy threshold that splits the vocabulary in a set of collection-independent words and a set of collection-specific words.
The next section describes how to estimate this threshold as a hyperparameter for arbitrary datasets automatically.

\subsection{Estimation of the Entropy Threshold}
\label{sec:hyperparameter}
We rank every word according to its entropy-based termhood.
On the one hand, there are certainly collection-specific words, which occur many times in each document of one collection and never in any document of the other collection.
On the other hand, there are certainly collection-independent words, which occur often and with equal frequency in each document, independently of the collection.
In between, the uncertainty is highest for words with the fewest observations: hapax legomena.

We set the entropy threshold such that hapax legomena are closest to it.
This threshold corresponds to the highest uncertainty whether a word is collection-specific or collection-independent.
Even if a corpus contains no hapax legomenon at all, this threshold can be calculated.
Imagine, there is a hapax legomenon $w_{hl}$ that occurs exactly once in collection $c_1$ and never in collection $c_2$ in a dataset consisting of two collections.
After Laplace smoothing (adding one pseudocount occurrence to each collection), $w_{hl}$ has two occurrences in $c_1$ and one occurrence in $c_2$.
Thus, we can estimate $P(c_1|w_{hl})=\frac{2}{3}$ and $P(c_2|w_{hl})=\frac{1}{3}$.
The entropy $H_{w_{hl}}$ follows as:
\begin{align*}
&H_{w_{hl}} = \displaystyle\sum_{\substack{i\in\{1,2\}}} -P(c_i|w_{hl})\cdot log_2 P(c_i|w_{hl})
\end{align*}

For an arbitrary number of collections $n$, collections $c_1,c_2,\ldots,c_n$, $P(c_i|w_{hl})=\frac{2}{n+1}$ for $i=1$, and $P(c_i|w_{hl})=\frac{1}{n+1}$ for all integers $i$, $1<i\leq n$, this formula generalizes to:
\begin{align*}
&H_{w_{hl}} = \displaystyle\sum_{i\in\{1,2,\ldots,n\}} -P(c_i|w_{hl})\cdot log_2 P(c_i|w_{hl})
\end{align*}

According to Zipf's law for natural language corpora, a word's frequency is inversely proportional to its rank in frequency.
Thus, most words in a corpus occur exactly once and by definition are hapax legomena, having entropy value $H_{w_{hl}}$.
As a consequence, we can expect this value to be the most frequent entropy value in a dataset.
We show that this holds in our four evaluation datasets in Section~\ref{sec:clasification}.

If the entropy threshold is set below this most frequent value, all hapax legomena are collection-independent.
If the threshold is set above this value, all hapax legomena are collection-specific.
As a consequence, a small change of the entropy threshold results in a large shift of the proportion of collection-specific words.
Notice that Laplace smoothing has been applied already.
Otherwise, a word that occurs exactly once would be collection-specific with certainty 100\%.
By applying Laplace smoothing, certainty for being collection-specific ranges from 0 (inclusive) to 1 (exclusive).

Because hapax legomena have been observed once in one collection and never in any other collection, we consider them as collection-specific words.
Topic representations list only the most frequent words per topic and therefore, hapax legomena typically do not appear in these representations.

$H_{w_{hl}}$ splits the vocabulary into two sets according to the entropy ranking, thus defining a set of collection-specific and a set of collection-independent words.
From the sizes of these sets and word frequency observations in the training dataset, we estimate the hyperparameter $\gamma$, which corresponds to the proportion of collection-specific words in the dataset.
$\gamma$ is estimated by the sum of occurrences of all collection-specific words divided by the total number of occurrences of words in the dataset. 

For the estimation of word distributions and topic distributions, we use Gibbs sampling based on the equations described in the original ccLDA paper~\cite{paul2009crosscultural}.
Our implementation of the entropy-based cross-collection model is open-sourced online\footnote{\url{https://hpi.de/naumann/projects/web-science/cross-collection-text-mining/entropy-based-topic-modeling.html}}.

\subsection{Further Considerations}
\paragraph{Arbitrary Number of Collections}
For two collections, the definition of collection-specific words is intuitive: these words occur most likely in one collection and unlikely in the other.
For more than two collections, for example three, this definition is less intuitive:
If a word occurs once in each of two collections but not in the third collection, is this word collection-specific?
We compute the entropy of a hapax legomenon as the entropy threshold, which separates collection-specific and collection-independent words.
Based on this threshold and our definition of collection-specificity, we consider words that occur once in two collections but not in the third collection as collection-independent, because their entropy is higher than the entropy of a hapax legomenon.
The broader definition of collection-specificity is justified, because it considers a word as collection-specific if its frequency distribution across collections differs from a uniform distribution with larger extent than a hapax legomenon's distribution.
We evaluate our model on datasets with two and with three collections in Section~\ref{sec:evaluation:results} and show examples for collection-specific words.

\paragraph{Multi-Lingual Corpora}
Our entropy-based topic model is suited for multi-lingual corpora as well.
If a corpus contains collections of two different languages, the linguistic differences are especially strong.
In this case, collection-independent words are words that occur in documents of either language, such as named entities or loanwords.
Collection-specific words are language-specific words that have no correspondence with the same spelling in the other language.
Our topic model provides topic-wise language-specific and language-independent words, which makes the results of our work interesting for linguists and machine translation.
In Section~\ref{sec:evaluation:results}, we evaluate our model's multi-lingual capabilities prototypically on English, French, and German Wikipedia articles.

\paragraph{Multi-Word Phrases}
To reduce semantic ambiguity of single words, we tokenize multi-word phrases during text pre-processing.
For an automatic phrase segmentation, we use the AutoPhrase algorithm of Liu et al.~\cite{shang2017automated}.
Therefore, we process each dataset with their phrase mining approach and segment multi-word phrases as single tokens automatically.
No labeled input phrases are required.

\paragraph{Application to ATE}
To use our model for the automatic extraction of domain terms, domains correspond to collections.
As a result after training our topic model, the most frequent top-k words from the collection-specific word distribution correspond to domain terms.
In the field of ATE, typical values for $k$ are 10 to 100~\cite{fedorenko2013automatic}.
An application to automatic term extraction would profit from multi-word phrases and their potential to reduce semantic ambiguity.

\section{Evaluation}
\label{sec:evaluation}
To compare ccLDA with the entropy-based model, we evaluate accuracy of document classification, topic coherence, and perplexity on datasets consisting of either two or three collections from different domains.

\subsection{Datasets}
\label{sec:evaluation:datasets}
We present four datasets, which differ in domain, language, number of documents, average document length, and number of collections.
Table~\ref{tab:datasets} gives an overview of the dataset sizes.

\begin{table}\centering
\caption{Collections with their respective number of documents $D$ and average number of words per document $W/D$ (after stop word removal).}
\ra{1.1}
\setlength{\tabcolsep}{3.8pt}
\begin{tabular}{@{}llrrrr@{}} \toprule
\textbf{Dataset} & \textbf{Collection} & $\mathbf{D}$ &$\mathbf{W/D}$ \\ \midrule
\multirow{2}{*}{Patents-Papers} & Patents &  14031 & 71\\ 
      & Papers &  16998 & 85\\ \cline{2-4}
    \multirow{2}{*}{Newspapers} & Guardian & 30774 & 310\\                            
            & Telegraph & 30749 & 273\\ \cline{2-4}  
    \multirow{3}{*}{Wikipedia} & English & 2927 & 324 \\
        & French & 2927 & 291 \\                          
            & German & 2927 & 302 \\ \cline{2-4}    
    \multirow{3}{*}{Traveler Forum} & India & 1432 & 199  \\                            
            & Singapore & 1179 & 187 \\ 
      & UK & 1580 & 288 \\
\bottomrule
\end{tabular}
\label{tab:datasets}
\end{table}
The patents-papers dataset focuses on the domain of computer science and contains abstracts of ACM papers and U.S. patents.
The patents\fnurl{https://www.uspto.gov/learning-and-resources/electronic-bulk-data-products} have been published by the United States Patent and Trademark Office (USPTO) between 2001 and 2016 and the papers\fnurl{https://aminer.org/citation} by Tang et al.~\cite{tang2008arnetminer}.
We consider only those patents that contain the term ``ACM'' in the citations.
To show that our approach generalizes to other document types, we evaluate our model on two large British newspapers: The Guardian and The Telegraph.
The dataset of articles published in the politics category between 2010 and 2015 forms the largest dataset in our evaluation.

The third dataset consists of Wikipedia articles that have an English, a French, and a German version. 
We crawled English, French, and German Wikipedia articles about movies produced between 2000 and 2016.
The length of such articles differs heavily, because the multi-lingual versions are no direct translations of each other.
Therefore, for each triple of an English, a French, and a German Wikipedia article, we calculate the minimum number of words and reduce all three articles to this text length.
If the minimum number of words is less than 300 (before stop word removal), the articles are discarded to ensure that the topic model learns only on articles of sufficient length.
Longer texts make sure that there are enough topic-specific words to learn a document's topics.

From the paper that proposes ccLDA, we reuse a dataset\fnurl{http://cmci.colorado.edu/~mpaul/downloads/ccdata.php} crawled from the online platform lonelyplanet.com~\cite{paul2009crosscultural}.
At this platform, there are three separate forums for Singapore, India, and the UK, each having thousands of threads.
Each document in Paul's dataset is the concatenation of all messages in one forum thread.

Table~\ref{table:entropythresholds} lists the four datasets with their natural number of topics, entropy thresholds (depending on the number of collections), and proportion of collection-specific words $\gamma$.
We determine the natural number of topics per dataset with a preliminary experiment following Arun et al.'s approach~\cite{arun2010finding}.
Their approach views LDA as a matrix factorization method and determines the natural number of topics for a given dataset by minimizing KL-Divergence of Singular value distributions.
The hyperparameter $\gamma$ is estimated as described in Section~\ref{sec:hyperparameter} and gives a first impression of the linguistic contrasts to expect in each dataset.
Not surprisingly, the proportion of collection-specific words is highest in the multi-lingual Wikipedia dataset, whereas it is lowest in newspaper articles.
Regarding the number of topics, patents and papers have the by far highest topical diversity.
\begin{table}\centering
\caption{Number of topics, entropy thresholds, and respective estimated hyperparameters $\gamma$ (proportion of collection-specific words).}
\ra{1.1}
\setlength{\tabcolsep}{3.8pt}
\begin{tabular}{@{}lrrr@{}} \toprule
\textbf{Dataset} & \textbf{\#Topics} & \textbf{Entropy} & $\mathbf{\gamma}$\\ \midrule
Patents-Papers & 260 & 0.918 & 0.632\\
Newspapers & 155 & 0.918 & 0.184\\
Wikipedia & 25 & 0.946 & 0.784\\
Traveler Forum & 100 & 0.946 & 0.742\\
\bottomrule
\end{tabular}
\label{table:entropythresholds}
\end{table}

\subsection{Experimental Setup}
The pre-processing consists of phrase segmentation, tokenization of multi-word phrases as single tokens\footnote{To ensure a fair comparison, we incorporate multi-word phrases also into the ccLDA baseline approach.}, and removal of stop words based on stop word lists.
The topic distribution priors, $\alpha$, are fixed and uniform except for one background topic.
According to Paul and Girju, updating $\alpha$ or other hyperparameters at runtime does not largely affect the sampling procedure~\cite{paul2009crosscultural}.
We set $\beta$ and $\delta$ to $0.01$ and $\gamma_0$ and $\gamma_1$ to $1.0$, corresponding to symmetric distributions with equal probability of occurrence of collection-specific and collection-independent words.
The Gibbs sampling runs for a burn-in period of 200 iterations.
10 samples, separated by lags of 10 iterations, are averaged for the final result.

After the sampling process, we measure per-topic coherence separately for the collection-independent word distribution $\varphi$ and each collection-specific word distribution $\sigma$.
Furthermore, we measure mixed topic coherence, language model perplexity, and document classification accuracy.
We split each dataset into 90\% training set and 10\% test set and use 10-fold cross-validation.

\subsection{Document Classification}

\begin{figure*}%
\begin{minipage}[b]{.25\textwidth}%
\centering%
\includegraphics[width=\textwidth]{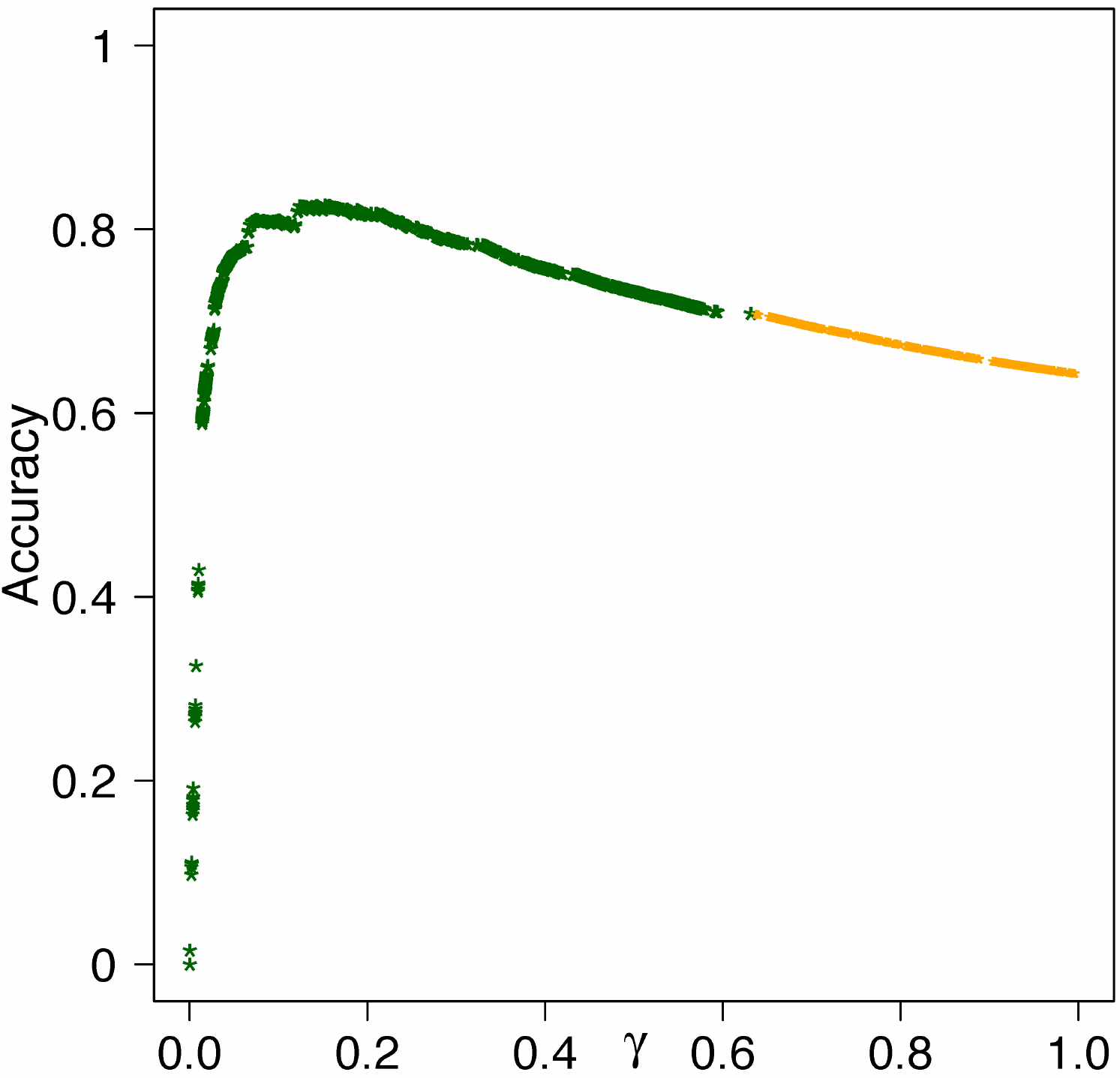}%
\subcaption{Patents-Papers}\label{fig:1d}%
\end{minipage}%
\begin{minipage}[b]{.25\textwidth}%
\centering%
\includegraphics[width=\textwidth]{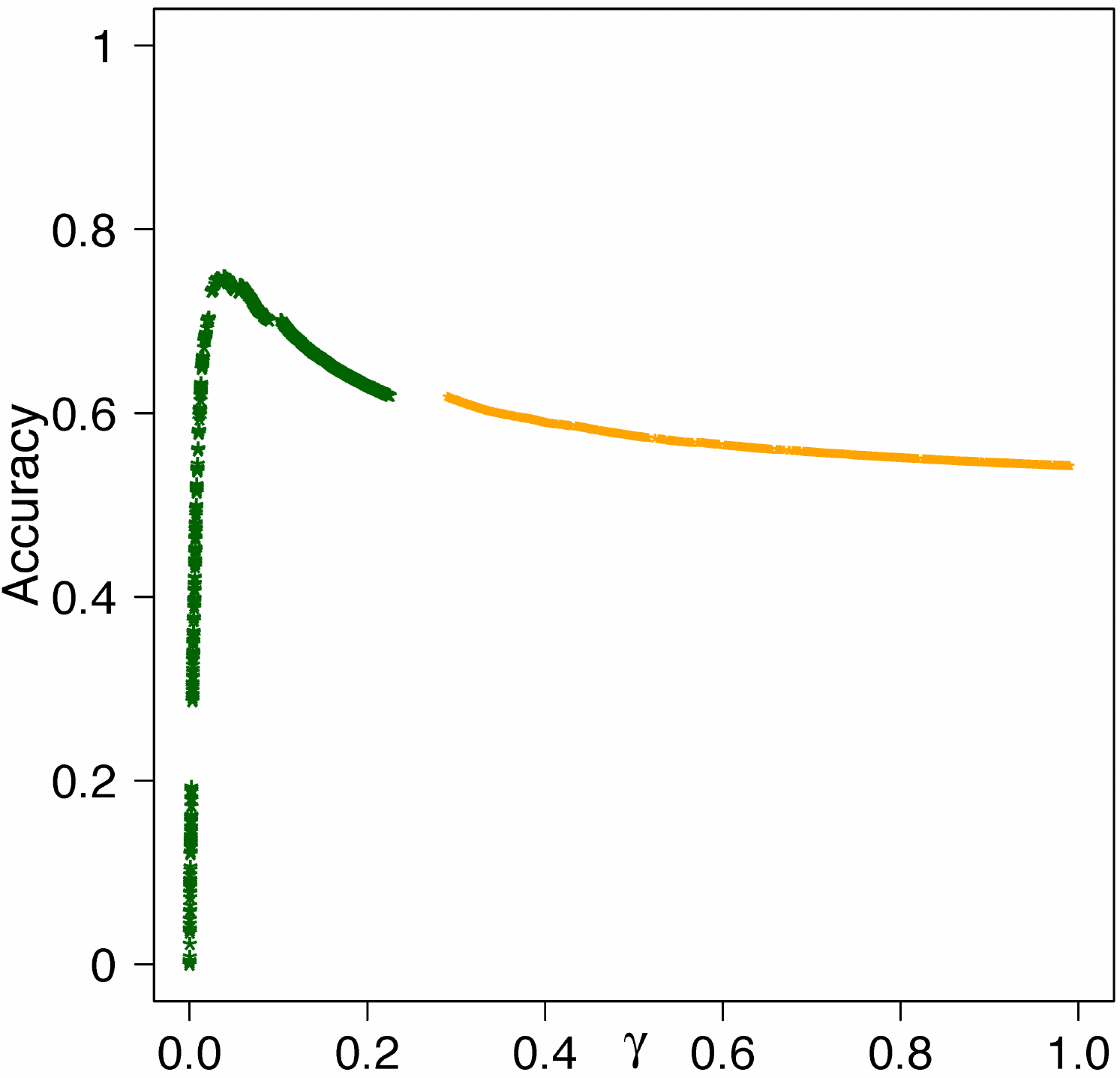}%
\subcaption{Newspapers}\label{fig:1b}%
\end{minipage}%
\begin{minipage}[b]{.25\textwidth}%
\centering%
\includegraphics[width=\textwidth]{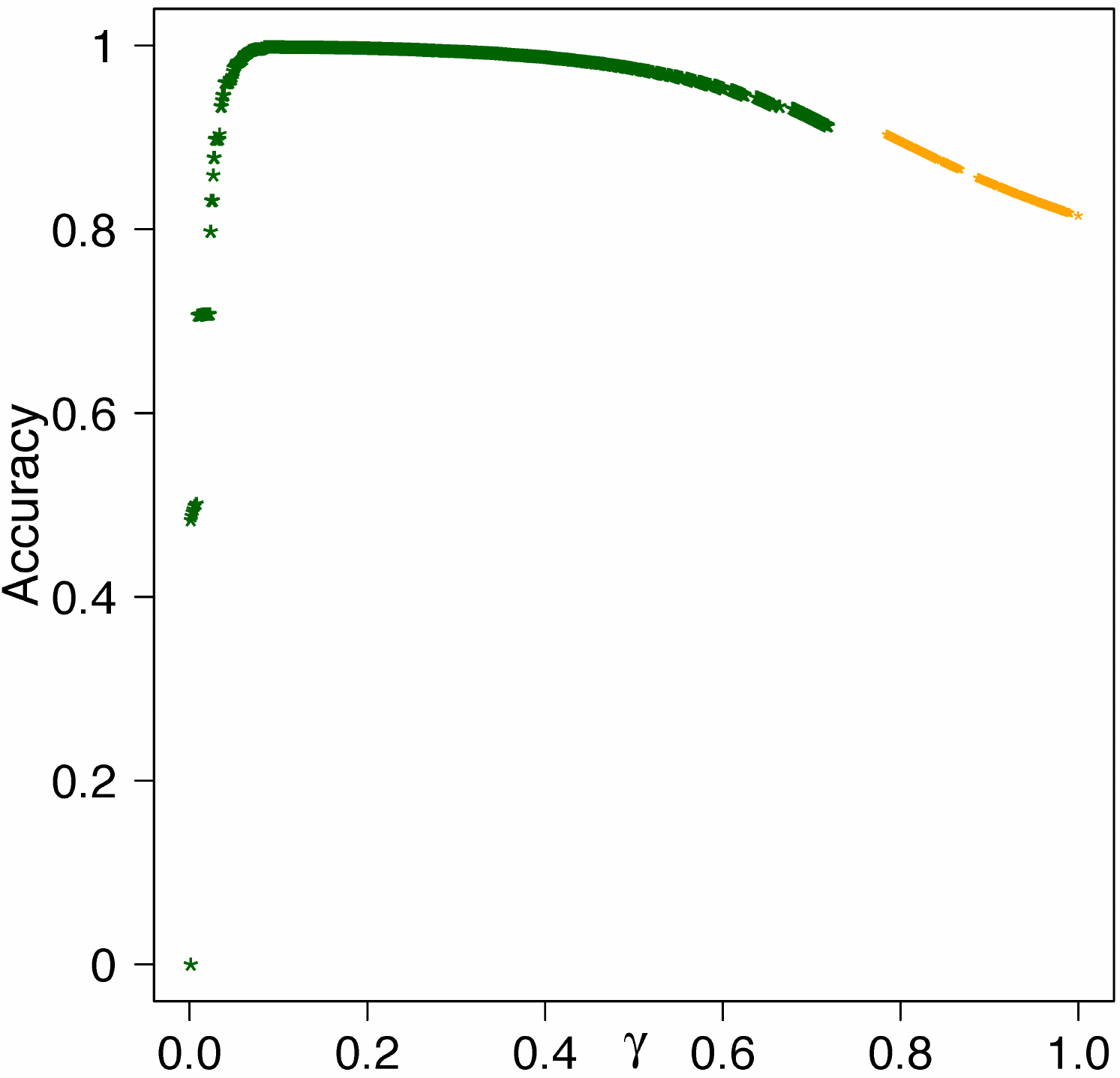}%
\subcaption{Wikipedia}\label{fig:1c}%
\end{minipage}%
\begin{minipage}[b]{.25\textwidth}%
\centering%
\includegraphics[width=\textwidth]{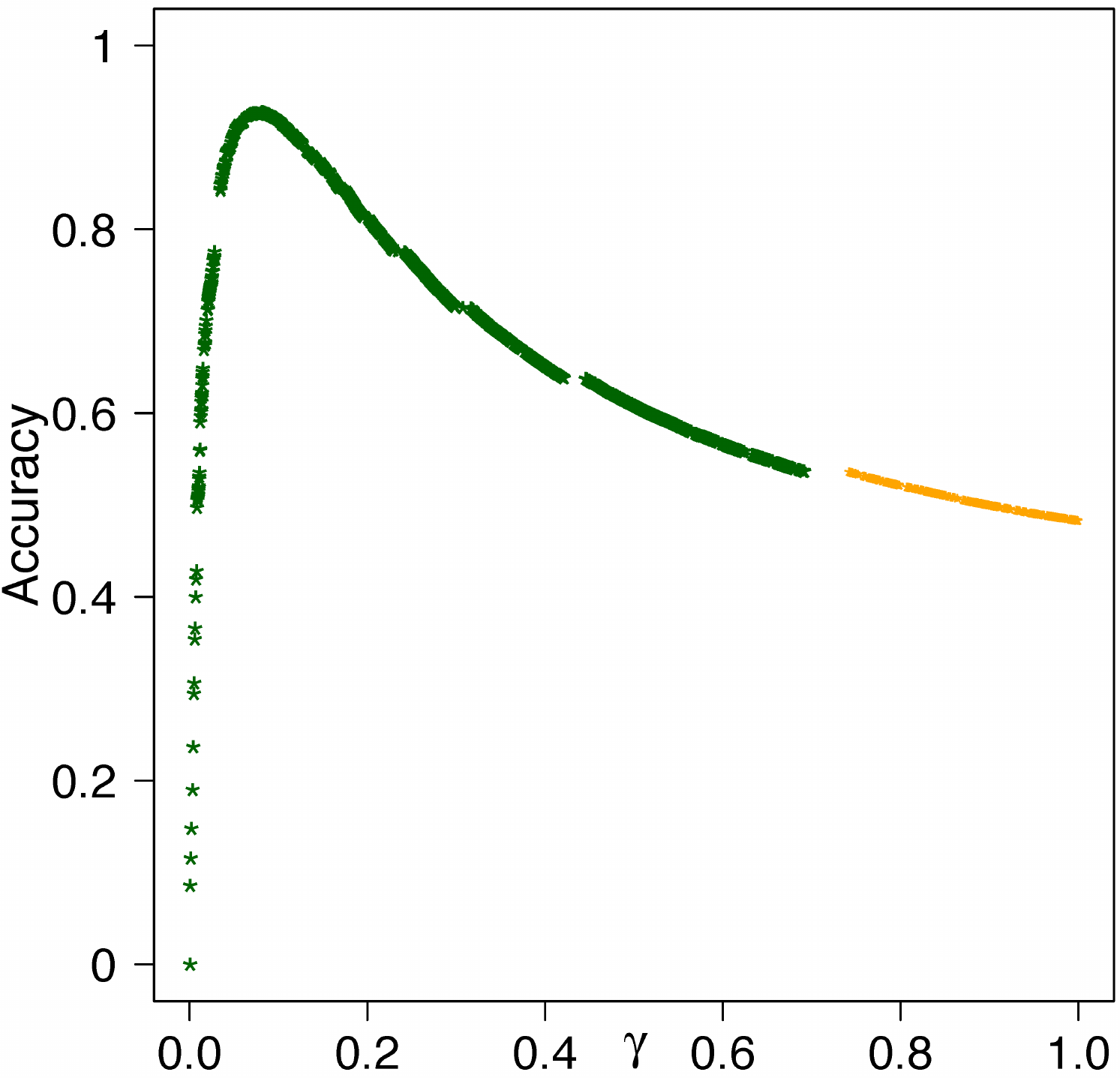}%
\subcaption{Traveler Forum}\label{fig:1a}%
\end{minipage}%
\caption{Document classification accuracy for all possible proportions of collection-specific words, $\gamma$. Collection-specific words in our entropy-based model are colored in green, collection-independent words in yellow.}
\label{fig:accuracies}%
\end{figure*}

\begin{table}\centering
\caption{Topic model comparison regarding accuracy (Acc), topic coherence (TC), and perplexity (Perpl). \\
\footnotesize{Asterisks denote statistically significant improvement as determined by a paired two-tailed t-test at 0.05 level.}}
\ra{1.1}
\setlength{\tabcolsep}{3.8pt}
\begin{tabular}{@{}lcccccc@{}} \toprule
& \multicolumn{3}{c}{\textbf{ccLDA}} & \multicolumn{3}{c}{\textbf{Entropy-Based}}\\
\textbf{Dataset} & \textbf{Acc} & \textbf{TC} & \textbf{Perpl} & \textbf{Acc} & \textbf{TC} & \textbf{Perpl} \\ \midrule
Patents-Papers & 0.61 & 0.404 & 827 & 0.70* & 0.412* & 797* \\
Newspapers & 0.54 & 0.436 & 4151 & 0.61* & 0.493* & 4042* \\
Wikipedia & 0.70 & - & 5150 & 0.92* & - & 4955* \\
Traveler Forum & 0.45 & 0.390 & 771 & 0.58* & 0.413* & 785 \\
\bottomrule
\end{tabular}
\label{table:results}
\end{table}

\label{sec:clasification}
The evaluation of the separation of collection-specific and collection-independent words is based on a document classification task.
For this task, each topic model predicts the collections of test documents given the documents' words.
Rather than only outputting the most likely collection per document, topic models assign a probability to each collection.
This probabilistic classification allows a more detailed evaluation regarding each topic model's degree of certainty.
The document classification accuracy corresponds to the probability assigned to the correct collection normalized by the sum of probabilities assigned to all collections.
We calculate the average accuracy across all test documents to obtain a document classification accuracy for each topic model.
This evaluation task allows for a comparison of ccLDA and the proposed topic model.\footnote{We do not consider document classification approaches as baselines, because the goal of this evaluation is to compare different cross-collection topic models.}

CcLDA calculates the probability $P(c|d)$ for collection $c$ given document $d$ as:
\[
    P(c|d) = \prod_{w\in d} \displaystyle\sum_{c} \mathbb{L}(w|\theta_d,c) 
\]
We use the word likelihood $\mathbb{L}(w|\theta_d,c)$ and compute the product of the likelihoods of all words in a document per collection $c$.
The classification accuracy for document $d$ is $\frac{P(c_{correct}|d)}{\sum_{c}P(c|d)}$.

Table~\ref{table:results} lists all evaluation results.
On the Wikipedia dataset, document classification accuracy of our entropy-based model is 31\% higher than ccLDA's accuracy.
On the other datasets, we achieve 29\% (traveler forum), 15\% (patents-papers), and 13\% (newspapers) higher accuracy.
Especially on datasets with stronger linguistic contrasts, such as the multi-lingual Wikipedia dataset, our approach outperforms ccLDA and provides a better separation of collection-specific and collection-independent words.
Comparing the different datasets, both approaches achieve the highest accuracy on the Wikipedia dataset.
This dataset contains many language-specific words.
It confirms that language classification of Wikipedia articles is easier than, for example, classification of newspaper articles as Guardian or Telegraph articles.

Figure~\ref{fig:accuracies} visualizes the influence of hyperparameter $\gamma$ on the document classification accuracy of our model.
The number of possible, meaningful hyperparameter settings is the number of distinct entropy values for words in the entire vocabulary. 
Thus, vocabulary size $V$ is an upper limit for the number of thresholds.
Each possible threshold corresponds to one hyperparameter setting and splits the vocabulary into two sets of collection-specific and collection-independent words.
For each possible threshold, we compute the topic model's document classification accuracy.
The larger the proportion of collection-specific words, the more words our model uses to classify a document.
The largest gap in each graph corresponds to hapax legomena.
We can either choose to consider all hapax legomena as collection-specific words or consider all hapax legomena as collection-independent words.
There is no choice in between, because all these words have the same entropy value.
The smaller gaps in the graph correspond to words that occur exactly twice (dis legomena), three times (tris legomena), and so on.
These words make up a large proportion of all words in the corpus, which is why their entropy value is more frequent.

Additionally, Figure~\ref{fig:accuracies} shows that we could achieve the highest document classification accuracy with $\gamma\approx0.1$.
However, such $\gamma$ considers words as collection-specific only if they improve document classification accuracy.
Incorrectly, the word's frequency distribution would not be considered.
For example, consider one word that occurs in every document of one collection and never in the other collection. 
This single, collection-specific word achieves optimal document classification accuracy.
Thus, all other words would be incorrectly considered collection-independent regardless of their frequency distribution.

\subsection{Topic Coherence}
The evaluation of topic coherence compares ccLDA and the entropy-based model with regard to their capability to cluster words by semantic similarity within one collection and across multiple collections.
Especially, we evaluate the capability to correctly align topics of different collections despite different word distributions.
For automatic evaluation of topic coherence, we use the Palmetto\fnurl{https://github.com/AKSW/Palmetto/} library, which implements the $C_V$ measure~\cite{Roder:2015:EST:2684822.2685324}.  

However, current topic coherence measures consider only single word distributions per topic and their topic representation in the form of top words.
Current measures cannot handle multiple word distributions per topic.
Therefore, we present an extension of topic coherence measures for cross-collection topic models, which we also published online.
Our approach considers topic representations (which are sets of top words) of the collection-independent word distribution $\varphi$ and each collection-specific word distribution $\sigma$.
Per topic, we use the union of these representations as a single topic representation across the collection-independent and each collection-specific topic-word distribution.
The coherence of this union can be measured with current topic coherence measures.
We call this measure ``mixed topic coherence'', because it mixes the word distributions and thereby allows to evaluate the topical alignment of the different word distributions.

However, measuring the topic coherence based on word co-occurrences has its limitations.
The $C_V$ measure assumes that words that never co-occur in the reference dataset are not topically coherent.
This assumption does not hold for datasets with strong linguistic contrast.
For example, words from collections with different languages do not co-occur although they can form a coherent topic.
Therefore, the $C_V$ measure cannot evaluate topic coherence of topic models trained on our multi-lingual dataset.
For all other datasets, we obtain a single coherence value per topic model by averaging all its topics' coherence values.
In the evaluation, ccLDA and the proposed entropy-based model form almost equally coherent topic representations with slightly higher topic coherence of our approach, especially on the newspapers dataset (13\% improvement).

\subsection{Perplexity}
With test set perplexity, we evaluate a topic model's ability to generalize from training data to unseen test data.
Perplexity focuses on the generative aspect of topic models to predict word probabilities for unseen documents.
Lower perplexity is better.

We use the ``fold-in'' method used by Paul et al.\ for the evaluation of ccLDA to learn topic probabilities $\theta$ of test documents.
``Fold-in'' means that the evaluation keeps all topic model parameters fixed as they have been learned on the training data. 
Gibbs sampling estimates only the per-document topic distributions on the test set.
This method has been introduced by Hofmann~\cite{hofmann1999probabilistic}.
Given the topic probabilities $\theta_d$ and the collection $c$ of a document $d$, the likelihood $\mathbb{L}$ of a word $w$ in $d$ is calculated as:
\[
\mathbb{L}(w|\theta_d,c) = \displaystyle\sum_{z}^{} P(z|\theta_d) [P(\neg x) P(w|z,\neg x)+ P(x) P(w|z,c,x)].
\]
In this formula, $x$ is a binary variable that denotes whether $w$ is collection-specific ($x$) or collection-independent ($\neg x$).
In ccLDA, this variable depends on collection $c$ and topic $z$ but not on the word itself. 
In contrast, our model estimates $x$ for each word in the vocabulary and thus learns the proportion of collection-specific words $\gamma$ per dataset.

$P(x)$ is the probability that a collection-specific word is sampled and is equivalent to $\gamma$ in our approach.
$P(w|z,x)$ is the probability that in particular word $w$ is sampled if a collection-specific word from topic $z$ is sampled.
The perplexity of a model $m$, with $M$ being the total number of words in all test documents, is calculated as:
\[
\mathbb{P}(m) = 2^{-\frac{1}{M}\prod_{w}\mathbb{L}(w|\theta_d,c)}.
\]
Both models achieve comparable perplexity with slight advantages of the entropy-based model, especially on the patents-papers dataset (4\% improvement).
On the traveler forum dataset, ccLDA achieves lower perplexity than the entropy-based model, although this difference is not significant.

\subsection{Example Topics}
\label{sec:evaluation:results}
As an example, Table~\ref{table:patentpapermodel} shows the top-5 words of each specific and independent word distribution our model learned on the patents-papers dataset.
For reasons of clarity, we limit the number of topics to 25.
Topic~0, the stop word topic, correctly identifies patent-specific stop words, such as ``apparatus'' and paper-specific stop words, such as ``approach''.
The word ``datum'', for example, is represented in the patent-specific and paper-specific word distributions, which means its collection-specific frequencies differ significantly and therefore cannot be combined in a single, collection-independent frequency.
Collection-independent words, such as ``method'' and ``system'', occur with similar frequency in both collections.
Another example, topic~5, deals with computer graphics (collection-independent top words) and our model shows that the focus of patents (``pixel'' or ``display'') is rather low-level, while papers focus on high-level algorithmic problems in the context of image processing (``algorithm'', ``detection'', ``recognition'').
Topic~2 is a security topic and its collection-specific words reveal that most documents about computer ``attacks'' are scientific papers, but no patents.

\setlength{\tabcolsep}{0.1em} 
{\renewcommand{\arraystretch}{0.9}
\begin{table*}
\caption{Entropy-based model with 25 topics for the patents-papers dataset.}
\centering\small
\begin{tabular}{|c|l|}
\hline
\textbf{ID} & Patent-Specific Topic Representation \hfill \textbf{Collection-Independent Topic Representation} \hfill Paper-Specific Topic Representation \\
\hline
0 &datum include computer apparatus determine\hfill \textbf{system method base provide process}\hfill model paper approach result datum \\
1 &task module tool script work\hfill \textbf{software agent group activity environment}\hfill student learning design computer learn\\
2 &message security user electronic access\_control\hfill \textbf{key secure signature authentication encryption}\hfill security scheme attack protocol access\_control\\
3 &circuit design test block programmable\hfill \textbf{logic integrate gate cell delay}\hfill test fault design circuit testing\\
4 &server client request user content\hfill \textbf{web\_service application network system platform}\hfill user mobile cloud architecture technology\\
5 &pixel display render unit frame\hfill \textbf{image color region camera method}\hfill feature object detection recognition algorithm\\
6 &user content template business knowledge\hfill \textbf{system digital provide media\_content base}\hfill science computer technology community conference\\
7 &database table result attribute schema\hfill \textbf{query search\_result retrieval xml relational}\hfill datum algorithm database optimization mining\\
8 &output input power second clock\hfill \textbf{signal frequency voltage supply current}\hfill power game low consumption circuit\\
9 &packet queue connection protocol switch\hfill \textbf{network route traffic communication flow}\hfill sensor\_nodes wireless protocol mobile\\
10 &channel terminal transmit receive wireless \hfill \textbf{communication transmission network receiver rate} \hfill channel performance antenna interference wireless\\
11 &partition cost constraint algorithm estimate\hfill \textbf{cell region method route threshold}\hfill water temperature thermal measurement study\\
12 &display user\_interface device graphical\hfill \textbf{virtual interactive provide environment visual}\hfill user\_interface interaction human device\\
13 &thread transaction memory request lock\hfill \textbf{resource system application scheduling distribute}\hfill memory parallel performance architecture processor\\
14 &device electronic mobile user location\hfill \textbf{customer network service product system}\hfill market risk patient health price\\
15 &document content input electronic feature\hfill \textbf{text word language character annotation}\hfill document gene semantic recognition feature\\
16 &element entry table hash classification\hfill \textbf{network rule set base machine}\hfill fuzzy classification neural\_network learning learn\\
17 &item segment list matrix determine\hfill \textbf{function method vector point space}\hfill problem matrix equation solution linear\\
18 &program execution instruction execute compile\hfill \textbf{source\_code language class logic type}\hfill program programming specification formal semantics\\
19 &object reference interface garbage pointer\hfill \textbf{management system process policy change}\hfill business knowledge project organization development\\
20 &node message graph plurality leaf\hfill \textbf{tree path number set time}\hfill algorithm problem graph bound complexity\\
21 &stream datum audio frame content\hfill \textbf{video filter signal code media}\hfill estimation algorithm noise propose scheme\\
22 &storage memory datum device block\hfill \textbf{system control controller volume primary}\hfill robot vehicle design simulation robotic\\
23 &instruction cache processor memory register\hfill \textbf{address window bit trace set}\hfill book guide include server learn\\
24 &model object threedimensional modeling polygon\hfill \textbf{surface point shape mesh flow}\hfill simulation finite\_element fluid property\\
\hline
\end{tabular}
\label{table:patentpapermodel}
\end{table*}
}

The original ccLDA is based on word co-occurrences only, which is particularly problematic if documents across collections share only few words, since they are then considered unlikely to belong to the same topic.
English, French, and German Wikipedia movie articles, for example, share less than $10\%$ of their words.
For comparison, the Guardian and the Telegraph articles share $37\%$.
As a consequence, ccLDA exhibits poor topic coherence and topic alignment across multi-lingual collections.

Table~\ref{table:comparewikipediamultilanguage} compares two Indian movie topics learned by the entropy-based model and ccLDA\@.
Both models are capable to assign English, French, and German words to respective collection-specific word distributions.
However, ccLDA considers ``khan'' to be the most frequent word of the collection-independent distribution and the English-specific distribution.
This mistake reveals ccLDA's deficiency to separate collection-specific and collection-independent words correctly.
In contrast, the entropy-based model guarantees that words are either collection-specific or collection-independent.
The word ``bollywood'' is not collection-independent, because its frequency is not uniformly distributed across all collections.
Instead, the word occurs more frequently in French articles with this topic (1.1\%) than in English articles (0.7\%).
Table~\ref{table:comparewikipediamultilanguage} further indicates that ccLDA does not align topics correctly:
The German-specific word distribution falsely contains sports related words and some words are not even German.

\setlength{\tabcolsep}{0.1em} 
{\renewcommand{\arraystretch}{0.9}
\begin{table}
\caption{Example topics from the Wikipedia dataset.}
\centering
\begin{tabular}{|c|c|c||c|c|c|}
\hline
\multicolumn{3}{|c||}{\textbf{Entropy-Based Model}} & \multicolumn{3}{c|}{\textbf{ccLDA}} \\ \hline
\multicolumn{3}{|c||}{khan judy arjun kabir} & \multicolumn{3}{c|}{khan hitch bragg} \\ 
\multicolumn{3}{|c||}{singh maya ali} & \multicolumn{3}{c|}{poojaclu cole vidya} \\
\multicolumn{3}{|c||}{estevez vidya hindus} & \multicolumn{3}{c|}{karan maya anjali} \\ 
\hline
\textbf{English} &\textbf{French} &\textbf{German} & \textbf{English} & \textbf{French} & \textbf{German} \\ \hline
indian & inde& adam& khan& mariage& chucky\\
india & indien& indien& indian & raj& team \\
 award& portail& hochzeit& india&  carter& carter\\
 bollywood& bollywood& liebe& film& inde & spieler\\
 love& composee&  verliebt& love & kapoor & cole\\
 kapoor& musique& tagebuch& award& ronan & spiel\\
 mitty& kapoor& brahm& raj& sam&  ella\\
 raj& interpretee & hideko& family& amoureux& duke\\
 english& technique& keat& shah& relation & kannibalen\\
 role& indienne& tiger& kapoor& pierre & jennifer \\
\hline
\end{tabular}
\label{table:comparewikipediamultilanguage}
\end{table}
}

Table~\ref{table:comparenewspaper} lists two topics about the military conflict in Syria and Iraq learned by ccLDA and the entropy-based model on the newspapers dataset.
The entropy-based model provides a better separation of collection-specific and collection-independent words.
For example, the entropy-based model assigns ``international'' to both collection-specific word distributions, because the word occurs with significantly different frequencies in both collections, which also holds for ``iraq'', ``syrian'', and ``assad''.
In both collection-specific word distributions, ``assad'' is the seventh most frequent word, but the difference of relative frequencies (1.2\% compared to 1.4\%) is comparably large so that our model represents the word with two collection-specific frequencies instead of a collection-independent frequency.
In contrast, ccLDA assigns ``britain'' to the collection-independent and the Telegraph-specific word distribution.
Representing ``britain'' in a collection-independent and a collection-specific distribution at the same time is inappropriate.
The entropy-based model reveals that the Telegraph and The Guardian differ in usage of the words ``isil'' and ``isis'' and that the politician William Hague is mentioned more frequently in The Telegraph for this topic.

\setlength{\tabcolsep}{0.1em} 
{\renewcommand{\arraystretch}{0.9}
\begin{table}
\caption{Example topics from the newspapers dataset.}
\centering
\begin{tabular}{|c|c||c|c|}
\hline
\multicolumn{2}{|c||}{\textbf{Entropy-Based Model}} & \multicolumn{2}{c|}{\textbf{ccLDA}} \\ \hline
\multicolumn{2}{|c||}{syria britain support country}  & \multicolumn{2}{c|}{russia country military attack} \\ 
\multicolumn{2}{|c||}{libya military force attack} & \multicolumn{2}{c|}{britain syria support} \\
\multicolumn{2}{|c||}{regime government british}  & \multicolumn{2}{c|}{russian world international} \\
\hline
 \textbf{Telegraph} &\textbf{Guardian} & \textbf{Telegraph}&\textbf{Guardian}  \\ \hline
  isil&iran & britain &force\\
  mr\_hague& iraq&  isil&state\\
  syrian& syrian&  \hspace{0.3cm}president\hspace{0.3cm} &president\\
  iraq& isis&  force&group\\
  air\_strikes&iraqi &  state&leader\\
  refugee& international&  america&power\\
  assad& assad&  british&cameron\\
 international&western &  leader&situation\\
  oil& civilian&  nation&america\\
  stop\_the\_war& humanitarian&  terrorist&official\\
\hline
\end{tabular}
\label{table:comparenewspaper}
\end{table}
}

Table~\ref{table:compareblogposts} compares transportation topics from the traveler forum corpus.
In the entropy-based model, the word ``luggage'' occurs in the India-specific and the Singapore-specific topic representation, because it is very frequent in this topic in both collection.
However, it is not collection-independent because its frequency distribution differs significantly from a uniform distribution with probabilities 0.018 in $C_{India}$, 0.029 in $C_{Singapore}$, and less than 0.01 in $C_{UK}$.
In ccLDA, the word ``time'' is falsely collection-specific \emph{and} collection-independent, occurring in the UK-specific, the Singapore-specific, and the collection-independent topic representation.

\setlength{\tabcolsep}{0.1em} 
{\renewcommand{\arraystretch}{0.9}
\begin{table}
\caption{Example topics from the traveler forum dataset.}
\centering
\begin{tabular}{|c|c|c||c|c|c|}
\hline
\multicolumn{3}{|c||}{\textbf{Entropy-Based Model}} & \multicolumn{3}{c|}{\textbf{ccLDA}} \\ \hline
\multicolumn{3}{|c||}{flight airport time hour check} & \multicolumn{3}{c|}{flight fly check hour airline} \\ 
\multicolumn{3}{|c||}{fly arrive book leave airline} & \multicolumn{3}{c|}{airport arrive time luggage bag} \\ \hline
\textbf{UK} & \textbf{India} & \textbf{Singapore} & \textbf{UK} & \textbf{India} & \textbf{Singapore} \\ \hline
heathrow &license &taxi &ticket &mumbaus &changi \\
train &\footnotesize{chandigarh} &changi& train &domestic &terminal\\
eurostar &luggage& terminal& book& mumbai& time\\
london& car &luggage &time& bangkok& luggage\\
ticket& bike &transit &allow &\footnotesize{international}& transit\\
frills &drive & \footnotesize{changi\_airport} &eurostar& \footnotesize{trivandrum} &\footnotesize{the\_airport}\\
paris &reliable &\footnotesize{the\_airport} &london& jet &\footnotesize{transit\_hotel}\\
\footnotesize{global\_guide}& valid &\footnotesize{immigration} &\footnotesize{connection}& air &free\\
\footnotesize{paddington} &storage &\footnotesize{transit\_hotel} &\footnotesize{global\_guide} &arrival &\footnotesize{changi\_airport}\\
option &enfield &shuttle &travel& direct& \footnotesize{immigration}\\
\hline
\end{tabular}
\label{table:compareblogposts}
\end{table}
}

\section{Conclusions and Future Work}
\label{sec:conclusion}
We presented a probabilistic, entropy-based topic model for multiple domain-specific text collections. 
Our approach incorporates multi-word phrases and is the first topic model that precisely distinguishes collection-specific and collection-independent words.
To compare our results to the state-of-the-art in cross-collection topic modeling, ccLDA, we evaluated document classification accuracy, topic coherence, and language model perplexity. 
We extended topic coherence measures from single-collection to cross-collection topic models and therefore allow to consider the alignment of word distributions in the evaluation.

Experiment results demonstrate the robustness of the proposed approach on a variety of text collections across different domains.
Our model outperforms ccLDA on all three quality measures on four evaluated datasets of two and three collections. 
Furthermore, our model provides superior topic representations due to its clear-cut separation of collection-specific and collection-independent words.
This separation is achieved by splitting the vocabulary into two sets of collection-specific and collection-independent words according to an entropy-based estimation of each word's termhood.
Whereas words with similar frequency across all collections are considered collection-independent, words with significantly different frequency per collection are considered collection-specific.

Possible applications are bias detection in newspaper articles or the revelation of regional and cultural differences in traveler forum posts.
Furthermore, topic-based search becomes possible in patents and scientific papers with collection-specific vocabulary and in multi-lingual document collections, such as Wikipedia.
While collection-specific word distributions reveal linguistic contrasts, collection-independent word distributions bridge the gap between collections by generalizing from domain-specific language.

For future work, an interesting extension of our approach is to calculate word entropy not only per corpus but per topic.
This extension considers words that are collection-specific in one topic and collection-independent in another topic.
Furthermore, the number of collections in our evaluation is limited to two and three.
Whether this number has a crucial effect needs to be studied in more detail and besides entropy, alternative measures of skewness could be tested.
Another possible path is further clustering of collection-specific words per-topic.
Opinion words are one example from related work that focuses on specific groups of words.
A clustering of words according to their function, role, or semantic context in a text document would be even more interesting.

\balance
\bibliographystyle{ACM-Reference-Format}
\bibliography{relatedwork} 


\providecommand{\noopsort}[1]{}
\begin{thebibliography}{00}


\ifx \showCODEN    \undefined \def \showCODEN     #1{\unskip}     \fi
\ifx \showDOI      \undefined \def \showDOI       #1{{\tt DOI:}\penalty0{#1}\ }
  \fi
\ifx \showISBNx    \undefined \def \showISBNx     #1{\unskip}     \fi
\ifx \showISBNxiii \undefined \def \showISBNxiii  #1{\unskip}     \fi
\ifx \showISSN     \undefined \def \showISSN      #1{\unskip}     \fi
\ifx \showLCCN     \undefined \def \showLCCN      #1{\unskip}     \fi
\ifx \shownote     \undefined \def \shownote      #1{#1}          \fi
\ifx \showarticletitle \undefined \def \showarticletitle #1{#1}   \fi
\ifx \showURL      \undefined \def \showURL       #1{#1}          \fi
\providecommand\bibfield[2]{#2}
\providecommand\bibinfo[2]{#2}
\providecommand\natexlab[1]{#1}
\providecommand\showeprint[2][]{arXiv:#2}

\bibitem[\protect\citeauthoryear{Arun, Suresh, Veni~Madhavan, and
  Narasimha~Murthy}{Arun et~al\mbox{.}}{2010}]%
        {arun2010finding}
\bibfield{author}{\bibinfo{person}{Rajkumar Arun},
  \bibinfo{person}{Venkatasubramaniyan Suresh}, \bibinfo{person}{C
  Veni~Madhavan}, {and} \bibinfo{person}{M Narasimha~Murthy}.}
  \bibinfo{year}{2010}\natexlab{}.
\newblock \showarticletitle{On finding the natural number of topics with latent
  dirichlet allocation: Some observations}. In \bibinfo{booktitle}{{\em Proc.
  of the Pacific-Asia Conf. on Advances in Knowledge Discovery and Data Mining
  (PAKDD)}}. \bibinfo{publisher}{Springer}, \bibinfo{pages}{391--402}.
\newblock


\bibitem[\protect\citeauthoryear{Bao, Collier, and Datta}{Bao
  et~al\mbox{.}}{2013}]%
        {bao2013partially}
\bibfield{author}{\bibinfo{person}{Yang Bao}, \bibinfo{person}{Nigel Collier},
  {and} \bibinfo{person}{Anindya Datta}.} \bibinfo{year}{2013}\natexlab{}.
\newblock \showarticletitle{A partially supervised cross-collection topic model
  for cross-domain text classification}. In \bibinfo{booktitle}{{\em Proc. of
  the Int. Conf. on Information and Knowledge Management (CIKM)}}. ACM,
  \bibinfo{pages}{239--248}.
\newblock


\bibitem[\protect\citeauthoryear{Blei, Ng, and Jordan}{Blei
  et~al\mbox{.}}{2003}]%
        {blei2003latent}
\bibfield{author}{\bibinfo{person}{David~M Blei}, \bibinfo{person}{Andrew~Y
  Ng}, {and} \bibinfo{person}{Michael~I Jordan}.}
  \bibinfo{year}{2003}\natexlab{}.
\newblock \showarticletitle{Latent dirichlet allocation}.
\newblock \bibinfo{journal}{{\em Journal of Machine Learning Research\/}}
  \bibinfo{volume}{3} (\bibinfo{year}{2003}), \bibinfo{pages}{993--1022}.
\newblock


\bibitem[\protect\citeauthoryear{Chang}{Chang}{2005}]%
        {chang2005domain}
\bibfield{author}{\bibinfo{person}{Jing-Shin Chang}.}
  \bibinfo{year}{2005}\natexlab{}.
\newblock \showarticletitle{Domain specific word extraction from hierarchical
  Web documents: A first step toward building lexicon trees from Web corpora}.
  In \bibinfo{booktitle}{{\em Proc. of the SIGHAN Workshop on Chinese Language
  Processing}}. ACL, \bibinfo{pages}{64--71}.
\newblock


\bibitem[\protect\citeauthoryear{Chen, Buntine, Ding, Xie, and Du}{Chen
  et~al\mbox{.}}{2015}]%
        {chen2015differential}
\bibfield{author}{\bibinfo{person}{Changyou Chen}, \bibinfo{person}{Wray
  Buntine}, \bibinfo{person}{Nan Ding}, \bibinfo{person}{Lexing Xie}, {and}
  \bibinfo{person}{Lan Du}.} \bibinfo{year}{2015}\natexlab{}.
\newblock \showarticletitle{Differential topic models}. In
  \bibinfo{booktitle}{{\em Transactions on Pattern Analysis and Machine
  Intelligence}}. IEEE, \bibinfo{pages}{230--242}.
\newblock


\bibitem[\protect\citeauthoryear{Eisenstein, Ahmed, and Xing}{Eisenstein
  et~al\mbox{.}}{2011}]%
        {eisenstein2011sparse}
\bibfield{author}{\bibinfo{person}{Jacob Eisenstein}, \bibinfo{person}{Amr
  Ahmed}, {and} \bibinfo{person}{Eric~P Xing}.}
  \bibinfo{year}{2011}\natexlab{}.
\newblock \showarticletitle{Sparse additive generative models of text}. In
  \bibinfo{booktitle}{{\em Proc. of the Int. Conf. on Machine Learning
  (ICML)}}. ACM, \bibinfo{pages}{1041--1048}.
\newblock


\bibitem[\protect\citeauthoryear{El-Kishky, Song, Wang, Voss, and
  Han}{El-Kishky et~al\mbox{.}}{2014}]%
        {El-Kishky:2014:STP:2735508.2735519}
\bibfield{author}{\bibinfo{person}{Ahmed El-Kishky}, \bibinfo{person}{Yanglei
  Song}, \bibinfo{person}{Chi Wang}, \bibinfo{person}{Clare~R. Voss}, {and}
  \bibinfo{person}{Jiawei Han}.} \bibinfo{year}{2014}\natexlab{}.
\newblock \showarticletitle{Scalable Topical Phrase Mining from Text Corpora}.
\newblock \bibinfo{journal}{{\em Proc. of the VLDB Endowment\/}}
  \bibinfo{volume}{8}, \bibinfo{number}{3} (\bibinfo{date}{Nov.}
  \bibinfo{year}{2014}), \bibinfo{pages}{305--316}.
\newblock


\bibitem[\protect\citeauthoryear{Fang, Si, Somasundaram, and Yu}{Fang
  et~al\mbox{.}}{2012}]%
        {fang2012mining}
\bibfield{author}{\bibinfo{person}{Yi Fang}, \bibinfo{person}{Luo Si},
  \bibinfo{person}{Naveen Somasundaram}, {and} \bibinfo{person}{Zhengtao Yu}.}
  \bibinfo{year}{2012}\natexlab{}.
\newblock \showarticletitle{Mining contrastive opinions on political texts
  using cross-perspective topic model}. In \bibinfo{booktitle}{{\em Proc. of
  the Int. Conf. on Web Search and Data Mining (WSDM)}}. ACM,
  \bibinfo{pages}{63--72}.
\newblock


\bibitem[\protect\citeauthoryear{Fedorenko, Astrakhantsev, and
  Turdakov}{Fedorenko et~al\mbox{.}}{2013}]%
        {fedorenko2013automatic}
\bibfield{author}{\bibinfo{person}{Denis Fedorenko}, \bibinfo{person}{Nikita
  Astrakhantsev}, {and} \bibinfo{person}{Denis Turdakov}.}
  \bibinfo{year}{2013}\natexlab{}.
\newblock \showarticletitle{Automatic recognition of domain-specific terms: an
  experimental evaluation.}, In \bibinfo{booktitle}{Proc. of the Spring
  Researchers Colloquium on Databases and Information Systems}.
  \bibinfo{journal}{{\em SYRCoDIS\/}} (\bibinfo{year}{2013}),
  \bibinfo{pages}{15--23}.
\newblock


\bibitem[\protect\citeauthoryear{Gao, Tang, Zhang, Jiang, Wu, and Zhuang}{Gao
  et~al\mbox{.}}{2012}]%
        {gao2012supervised}
\bibfield{author}{\bibinfo{person}{Haidong Gao}, \bibinfo{person}{Siliang
  Tang}, \bibinfo{person}{Yin Zhang}, \bibinfo{person}{Dapeng Jiang},
  \bibinfo{person}{Fei Wu}, {and} \bibinfo{person}{Yueting Zhuang}.}
  \bibinfo{year}{2012}\natexlab{}.
\newblock \showarticletitle{Supervised cross-collection topic modeling}. In
  \bibinfo{booktitle}{{\em Proc. of the Int. Conf. on Multimedia (MM)}}. ACM,
  \bibinfo{pages}{957--960}.
\newblock


\bibitem[\protect\citeauthoryear{Griffiths and Steyvers}{Griffiths and
  Steyvers}{2004}]%
        {griffiths2004finding}
\bibfield{author}{\bibinfo{person}{Thomas~L Griffiths} {and}
  \bibinfo{person}{Mark Steyvers}.} \bibinfo{year}{2004}\natexlab{}.
\newblock \showarticletitle{Finding scientific topics}.
\newblock \bibinfo{journal}{{\em Proc. of the National Academy of Sciences
  (PNAS)\/}} \bibinfo{volume}{101}, \bibinfo{number}{suppl 1}
  (\bibinfo{year}{2004}), \bibinfo{pages}{5228--5235}.
\newblock


\bibitem[\protect\citeauthoryear{He}{He}{2016}]%
        {He:2016:ETP:3016100.3016316}
\bibfield{author}{\bibinfo{person}{Yulan He}.} \bibinfo{year}{2016}\natexlab{}.
\newblock \showarticletitle{Extracting Topical Phrases from Clinical
  Documents}. In \bibinfo{booktitle}{{\em Proc. of the Nat. Conf. on Artificial
  Intelligence (AAAI)}}. \bibinfo{publisher}{AAAI Press},
  \bibinfo{pages}{2957--2963}.
\newblock


\bibitem[\protect\citeauthoryear{Hofmann}{Hofmann}{1999}]%
        {hofmann1999probabilistic}
\bibfield{author}{\bibinfo{person}{Thomas Hofmann}.}
  \bibinfo{year}{1999}\natexlab{}.
\newblock \showarticletitle{Probabilistic latent semantic indexing}. In
  \bibinfo{booktitle}{{\em Proc. of the Int. Conf. on Research and Development
  in Information Retrieval (SIGIR)}}. ACM, \bibinfo{pages}{50--57}.
\newblock


\bibitem[\protect\citeauthoryear{Jameel and Lam}{Jameel and Lam}{2013}]%
        {Jameel:2013:UTS:2484028.2484062}
\bibfield{author}{\bibinfo{person}{Shoaib Jameel} {and} \bibinfo{person}{Wai
  Lam}.} \bibinfo{year}{2013}\natexlab{}.
\newblock \showarticletitle{An Unsupervised Topic Segmentation Model
  Incorporating Word Order}. In \bibinfo{booktitle}{{\em Proc. of the Int.
  Conf. on Research and Development in Information Retrieval (SIGIR)}}.
  \bibinfo{publisher}{ACM}, \bibinfo{pages}{203--212}.
\newblock
\showISBNx{978-1-4503-2034-4}


\bibitem[\protect\citeauthoryear{Kageura and Umino}{Kageura and Umino}{1996}]%
        {kageura1996methods}
\bibfield{author}{\bibinfo{person}{Kyo Kageura} {and} \bibinfo{person}{Bin
  Umino}.} \bibinfo{year}{1996}\natexlab{}.
\newblock \showarticletitle{Methods of automatic term recognition: A review}.
\newblock \bibinfo{journal}{{\em Terminology\/}} \bibinfo{volume}{3},
  \bibinfo{number}{2} (\bibinfo{year}{1996}), \bibinfo{pages}{259--289}.
\newblock


\bibitem[\protect\citeauthoryear{Kawamae}{Kawamae}{2014}]%
        {Kawamae:2014:SNT:2556195.2559895}
\bibfield{author}{\bibinfo{person}{Noriaki Kawamae}.}
  \bibinfo{year}{2014}\natexlab{}.
\newblock \showarticletitle{Supervised N-gram Topic Model}. In
  \bibinfo{booktitle}{{\em Proc. of the Int. Conf. on Web Search and Data
  Mining (WSDM)}}. \bibinfo{publisher}{ACM}, \bibinfo{pages}{473--482}.
\newblock


\bibitem[\protect\citeauthoryear{Kit}{Kit}{2002}]%
        {kit2002corpus}
\bibfield{author}{\bibinfo{person}{Chunyu Kit}.}
  \bibinfo{year}{2002}\natexlab{}.
\newblock \showarticletitle{Corpus tools for retrieving and deriving termhood
  evidence}. In \bibinfo{booktitle}{{\em Proc. of the East Asia Forum of
  Terminology}}. \bibinfo{pages}{69--80}.
\newblock


\bibitem[\protect\citeauthoryear{Krestel and Smyth}{Krestel and Smyth}{2013}]%
        {krestel2013recommending}
\bibfield{author}{\bibinfo{person}{Ralf Krestel} {and}
  \bibinfo{person}{Padhraic Smyth}.} \bibinfo{year}{2013}\natexlab{}.
\newblock \showarticletitle{Recommending patents based on latent topics}. In
  \bibinfo{booktitle}{{\em Proc. of the Conf. on Recommender Systems
  (RecSys)}}. ACM, \bibinfo{pages}{395--398}.
\newblock


\bibitem[\protect\citeauthoryear{Lau, Baldwin, and Newman}{Lau
  et~al\mbox{.}}{2013}]%
        {Lau:2013:CTM:2483969.2483972}
\bibfield{author}{\bibinfo{person}{Jey~Han Lau}, \bibinfo{person}{Timothy
  Baldwin}, {and} \bibinfo{person}{David Newman}.}
  \bibinfo{year}{2013}\natexlab{}.
\newblock \showarticletitle{On Collocations and Topic Models}.
\newblock \bibinfo{journal}{{\em ACM Transactions on Speech and Language
  Processing (TSLP)\/}} \bibinfo{volume}{10}, \bibinfo{number}{3}
  (\bibinfo{date}{July} \bibinfo{year}{2013}), \bibinfo{pages}{10:1--10:14}.
\newblock
\showISSN{1550-4875}


\bibitem[\protect\citeauthoryear{Li, Li, Song, Li, and Chang}{Li
  et~al\mbox{.}}{2013}]%
        {li2013novel}
\bibfield{author}{\bibinfo{person}{Sujian Li}, \bibinfo{person}{Jiwei Li},
  \bibinfo{person}{Tao Song}, \bibinfo{person}{Wenjie Li}, {and}
  \bibinfo{person}{Baobao Chang}.} \bibinfo{year}{2013}\natexlab{}.
\newblock \showarticletitle{A novel topic model for automatic term extraction}.
  In \bibinfo{booktitle}{{\em Proc. of the Int. Conf. on Research and
  Development in Information Retrieval (SIGIR)}}. ACM,
  \bibinfo{pages}{885--888}.
\newblock


\bibitem[\protect\citeauthoryear{Lindsey, Headden, and Stipicevic}{Lindsey
  et~al\mbox{.}}{2012}]%
        {Lindsey:2012:PTM:2390948.2390975}
\bibfield{author}{\bibinfo{person}{Robert~V. Lindsey},
  \bibinfo{person}{William~P. Headden, III}, {and} \bibinfo{person}{Michael~J.
  Stipicevic}.} \bibinfo{year}{2012}\natexlab{}.
\newblock \showarticletitle{A Phrase-discovering Topic Model Using Hierarchical
  Pitman-Yor Processes}. In \bibinfo{booktitle}{{\em Proc. of the Joint Conf.
  on Empirical Methods in Natural Language Processing and Computational Natural
  Language Learning (EMNLP-CoNLL)}}. \bibinfo{publisher}{ACL},
  \bibinfo{pages}{214--222}.
\newblock


\bibitem[\protect\citeauthoryear{Liu, Shang, Wang, Ren, and Han}{Liu
  et~al\mbox{.}}{2015}]%
        {liu2015mining}
\bibfield{author}{\bibinfo{person}{Jialu Liu}, \bibinfo{person}{Jingbo Shang},
  \bibinfo{person}{Chi Wang}, \bibinfo{person}{Xiang Ren}, {and}
  \bibinfo{person}{Jiawei Han}.} \bibinfo{year}{2015}\natexlab{}.
\newblock \showarticletitle{Mining quality phrases from massive text corpora}.
  In \bibinfo{booktitle}{{\em Proc. of the Int. Conf. on Management of Data
  (SIGMOD)}}. ACM, \bibinfo{pages}{1729--1744}.
\newblock


\bibitem[\protect\citeauthoryear{Paul and Girju}{Paul and Girju}{2009}]%
        {paul2009crosscultural}
\bibfield{author}{\bibinfo{person}{Michael Paul} {and} \bibinfo{person}{Roxana
  Girju}.} \bibinfo{year}{2009}\natexlab{}.
\newblock \showarticletitle{Cross-cultural analysis of blogs and forums with
  mixed-collection topic models}. In \bibinfo{booktitle}{{\em Proc. of the
  Conf. on Empirical Methods in Natural Language Processing (EMNLP)}}. ACL,
  \bibinfo{pages}{1408--1417}.
\newblock


\bibitem[\protect\citeauthoryear{Risch and Krestel}{Risch and Krestel}{2017}]%
        {risch2017what}
\bibfield{author}{\bibinfo{person}{Julian Risch} {and} \bibinfo{person}{Ralf
  Krestel}.} \bibinfo{year}{2017}\natexlab{}.
\newblock \showarticletitle{What Should I Cite? Cross-Collection Reference
  Recommendation of Patents and Papers}. In \bibinfo{booktitle}{{\em Proc. of
  the Int. Conf. on Theory and Practice of Digital Libraries (TPDL)}}.
  \bibinfo{publisher}{Springer}, \bibinfo{pages}{40--46}.
\newblock


\bibitem[\protect\citeauthoryear{R{\"o}der, Both, and Hinneburg}{R{\"o}der
  et~al\mbox{.}}{2015}]%
        {Roder:2015:EST:2684822.2685324}
\bibfield{author}{\bibinfo{person}{Michael R{\"o}der}, \bibinfo{person}{Andreas
  Both}, {and} \bibinfo{person}{Alexander Hinneburg}.}
  \bibinfo{year}{2015}\natexlab{}.
\newblock \showarticletitle{Exploring the space of topic coherence measures}.
  In \bibinfo{booktitle}{{\em Proc. of the Int. Conf. on Web Search and Data
  Mining (WSDM)}}. ACM, \bibinfo{pages}{399--408}.
\newblock


\bibitem[\protect\citeauthoryear{Shang, Liu, Jiang, Ren, Voss, and Han}{Shang
  et~al\mbox{.}}{2018}]%
        {shang2017automated}
\bibfield{author}{\bibinfo{person}{Jingbo Shang}, \bibinfo{person}{Jialu Liu},
  \bibinfo{person}{Meng Jiang}, \bibinfo{person}{Xiang Ren},
  \bibinfo{person}{Clare~R Voss}, {and} \bibinfo{person}{Jiawei Han}.}
  \bibinfo{year}{2018}\natexlab{}.
\newblock \showarticletitle{Automated phrase mining from massive text corpora}.
\newblock \bibinfo{journal}{{\em IEEE Transactions on Knowledge and Data
  Engineering\/}} (\bibinfo{year}{2018}).
\newblock


\bibitem[\protect\citeauthoryear{Tang, Zhang, Yao, Li, Zhang, and Su}{Tang
  et~al\mbox{.}}{2008}]%
        {tang2008arnetminer}
\bibfield{author}{\bibinfo{person}{Jie Tang}, \bibinfo{person}{Jing Zhang},
  \bibinfo{person}{Limin Yao}, \bibinfo{person}{Juanzi Li}, \bibinfo{person}{Li
  Zhang}, {and} \bibinfo{person}{Zhong Su}.} \bibinfo{year}{2008}\natexlab{}.
\newblock \showarticletitle{Arnetminer: Extraction and mining of academic
  social networks}. In \bibinfo{booktitle}{{\em Proc. of the Int. Conf. on
  Knowledge Discovery and Data Mining (SIGKDD)}}. ACM,
  \bibinfo{pages}{990--998}.
\newblock


\bibitem[\protect\citeauthoryear{Wang, McCallum, and Wei}{Wang
  et~al\mbox{.}}{2007}]%
        {wang2007topical}
\bibfield{author}{\bibinfo{person}{Xuerui Wang}, \bibinfo{person}{Andrew
  McCallum}, {and} \bibinfo{person}{Xing Wei}.}
  \bibinfo{year}{2007}\natexlab{}.
\newblock \showarticletitle{Topical n-grams: Phrase and topic discovery, with
  an application to information retrieval}. In \bibinfo{booktitle}{{\em Proc.
  of the Int. Conf. on Data Mining (ICDM)}}. IEEE, \bibinfo{pages}{697--702}.
\newblock


\bibitem[\protect\citeauthoryear{Wilson and Chew}{Wilson and Chew}{2010}]%
        {wilson2010term}
\bibfield{author}{\bibinfo{person}{Andrew~T Wilson} {and}
  \bibinfo{person}{Peter~A Chew}.} \bibinfo{year}{2010}\natexlab{}.
\newblock \showarticletitle{Term weighting schemes for latent dirichlet
  allocation}. In \bibinfo{booktitle}{{\em Proc. of the Conf. of the North
  American Chapter of the Association for Computational Linguistics (NAACL)}}.
  ACL, \bibinfo{pages}{465--473}.
\newblock


\bibitem[\protect\citeauthoryear{Yao, Mimno, and McCallum}{Yao
  et~al\mbox{.}}{2009}]%
        {yao2009efficient}
\bibfield{author}{\bibinfo{person}{Limin Yao}, \bibinfo{person}{David Mimno},
  {and} \bibinfo{person}{Andrew McCallum}.} \bibinfo{year}{2009}\natexlab{}.
\newblock \showarticletitle{Efficient methods for topic model inference on
  streaming document collections}. In \bibinfo{booktitle}{{\em Proc. of the
  Int. Conf. on Knowledge Discovery and Data Mining (SIGKDD)}}. ACM,
  \bibinfo{pages}{937--946}.
\newblock


\bibitem[\protect\citeauthoryear{Zhai, Velivelli, and Yu}{Zhai
  et~al\mbox{.}}{2004}]%
        {zhai2004cross}
\bibfield{author}{\bibinfo{person}{ChengXiang Zhai}, \bibinfo{person}{Atulya
  Velivelli}, {and} \bibinfo{person}{Bei Yu}.} \bibinfo{year}{2004}\natexlab{}.
\newblock \showarticletitle{A cross-collection mixture model for comparative
  text mining}. In \bibinfo{booktitle}{{\em Proc. of the Int. Conf. on
  Knowledge Discovery and Data Mining (SIGKDD)}}. ACM,
  \bibinfo{pages}{743--748}.
\newblock


\bibitem[\protect\citeauthoryear{Zhang, Mei, and Zhai}{Zhang
  et~al\mbox{.}}{2010}]%
        {zhang2010cross}
\bibfield{author}{\bibinfo{person}{Duo Zhang}, \bibinfo{person}{Qiaozhu Mei},
  {and} \bibinfo{person}{ChengXiang Zhai}.} \bibinfo{year}{2010}\natexlab{}.
\newblock \showarticletitle{Cross-lingual latent topic extraction}. In
  \bibinfo{booktitle}{{\em Proc. of the Annual Meeting on Association for
  Computational Linguistics (ACL)}}. ACL, \bibinfo{pages}{1128--1137}.
\newblock


\bibitem[\protect\citeauthoryear{Zhang, Gerow, Altosaar, Evans, and So}{Zhang
  et~al\mbox{.}}{2015}]%
        {zhang2015fast}
\bibfield{author}{\bibinfo{person}{Jingwei Zhang}, \bibinfo{person}{Aaron
  Gerow}, \bibinfo{person}{Jaan Altosaar}, \bibinfo{person}{James Evans}, {and}
  \bibinfo{person}{Richard~Jean So}.} \bibinfo{year}{2015}\natexlab{}.
\newblock \showarticletitle{Fast, flexible models for discovering topic
  correlation across weakly-related collections}. In \bibinfo{booktitle}{{\em
  Proc. of the Conf. on Empirical Methods in Natural Language Processing
  (EMNLP)}}. ACL, \bibinfo{pages}{1554--1564}.
\newblock


\bibitem[\protect\citeauthoryear{Zheng, Nie, Moriya, Inoue, Imada, Utsuro,
  Kawada, and Kando}{Zheng et~al\mbox{.}}{2014}]%
        {zheng2014comparative}
\bibfield{author}{\bibinfo{person}{Liyi Zheng}, \bibinfo{person}{Tian Nie},
  \bibinfo{person}{Ichiro Moriya}, \bibinfo{person}{Yusuke Inoue},
  \bibinfo{person}{Takakazu Imada}, \bibinfo{person}{Takehito Utsuro},
  \bibinfo{person}{Yasuhide Kawada}, {and} \bibinfo{person}{Noriko Kando}.}
  \bibinfo{year}{2014}\natexlab{}.
\newblock \showarticletitle{Comparative topic analysis of japanese and chinese
  bloggers}. In \bibinfo{booktitle}{{\em Proc. of the Int. Conf. on Advanced
  Information Networking and Applications Workshops (WAINA)}}. IEEE,
  \bibinfo{pages}{664--669}.
\newblock


\end{thebibliography}

\end{document}